\documentclass[
 reprint,
 amsmath,amssymb,
 aps,
 prl,
]{revtex4-1}

\newcommand{\ra}[1]{\renewcommand{\arraystretch}{#1}}

\usepackage{graphicx}
\usepackage{dcolumn}
\usepackage{bm}
\usepackage[hypertexnames=false]{hyperref}
\usepackage{color}
\usepackage{xcolor} 
\usepackage{titlesec}

\definecolor{blue(pigment)}{rgb}{0.2, 0.2, 0.6}

\begin{document}


\title{The role of advective inertia in active nematic turbulence}

\author{Colin-Marius Koch}
\author{Michael Wilczek}%
 \email{michael.wilczek@ds.mpg.de}
\affiliation{%
 Max Planck Institute for Dynamics and Self-Organization, Am Fa{\ss}berg 17, 37077 G\"ottingen, Germany and \\ Faculty of Physics, Georg-August-Universit\"at G\"ottingen, \\ Friedrich-Hund-Platz 1, 37077 G\"ottingen, Germany
}%

\date{\today}

\begin{abstract}
Suspensions of active agents with nematic interactions exhibit complex spatio-temporal dynamics such as mesoscale turbulence. Since the Reynolds number of microscopic flows is very small on the scale of individual agents, inertial effects are typically excluded in continuum theories of active nematic turbulence. Whether active stresses can collectively excite inertial flows is currently unclear. To address this question, we investigate a two-dimensional continuum theory for active nematic turbulence. In particular, we compare mesoscale turbulence with and without the effects of advective inertia. We find that inertial effects can influence the flow already close to the onset of the turbulent state and, moreover, give rise to large-scale fluid motion for strong active driving. A detailed analysis of the kinetic energy budget reveals an energy transfer to large scales mediated by inertial advection. While this transfer is small in comparison to energy injection and dissipation, its effects accumulate over time. The inclusion of friction, which is typically present in experiments, can compensate for this effect. The findings suggest that the inclusion of inertia and friction may be necessary for dynamically consistent theories of active nematic turbulence.
\end{abstract}

\maketitle

Active matter on the microscale consists of motile agents, such as bacteria~\cite{dombrowski2004self,sokolov2007concentration,lushi2014fluid} and cells~\cite{kemkemer2000nematic,peruani2012collective,duclos2017topological}, filaments driven by motor proteins~\cite{ndlec1997self,surrey2001physical,schaller2010polar,sumino2012large,sanchez2012spontaneous}, motile algae~\cite{drescher2010direct,drescher2011fluid,von2020diffusive}, or colloids~\cite{bricard2013emergence,cao2019orientational}. Suspended densely in a liquid, they form so-called active fluids, in which the flow is driven on the scale of the agents~\cite{saintillan2018rheology}. Their collective behavior can lead to complex mesoscale phenomena, such as active turbulence, which is reminiscent of driven hydrodynamic flows and has been observed e.g.~in suspensions of bacteria~\cite{dombrowski2004self,wensink2012meso,dunkel2013fluid} and in microtubule kinesin mixtures~\cite{sanchez2012spontaneous,guillamat2016probing}. The latter case is an example of an active liquid crystal, for which continuum models have been adapted from liquid crystal theory to include active stresses that excite the flow field~\cite{simha2002hydrodynamic,thampi2016active,doostmohammadi2018active}.

An individual microscopic agent is subject to drag forces in the fluid, which are large compared to its inertial forces due to its small size, mass, and propulsion speed~\cite{purcell1977life,lauga2009hydrodynamics,elgeti2015physics}. As a consequence, the agent's dynamics are dominated by its self-propulsion and the viscous damping of the fluid. However, the collective motion that leads to active turbulence has been found to significantly exceed velocities found for individual agents~\cite{mendelson1999organized,dombrowski2004self}. This raises the question of whether the collective behavior of many active agents in principle can excite flows in which inertial effects become apparent. 

Here, we address this question with a detailed study on the impact of inertial forces on dense suspensions of active agents in the framework of an established two-dimensional continuum model of active nematic liquid crystals \cite{marenduzzo2007hydrodynamics,thampi2013velocity, marchetti2013hydrodynamics,giomi2011excitable}, which has been related to experimental results~\cite{guillamat2016control}. In this model for wet active matter, hydrodynamic interactions have either been taken into account using Stokes flow~\cite{hemingway2016correlation,doostmohammadi2018active,alert2020universal}, or unsteady Stokes flow~\cite{giomi2011excitable,giomi2012banding,giomi2014defect,giomi2015geometry}. Here, we additionally include the full Navier-Stokes dynamics to test whether active stresses can excite collective inertial flows. In particular, we explore under which conditions large-scale flow patterns like the ones observed in two-dimensional hydrodynamic turbulence can emerge~\cite{boffetta2010evidence,boffetta2012two,alexakis2018cascades}. To this end, we perform numerical simulations of this model with and without inertial advection and compare various flow statistics in both regimes.

To study the impact of advective inertia on the fluid flow of two-dimensional active nematic turbulence, we use the continuum equations established for a dense suspension of motile, aligning particles~\cite{giomi2015geometry}, which originate from the well-studied Beris-Edwards model of liquid crystal theory~\cite{beris1994thermodynamics,qian1998generalized,hemingway2016correlation, abels2016strong}. In essence, the equations couple the fluid flow to an order parameter field describing the nematic order. The fluid flow is described by the incompressible ($\nabla \cdot \boldsymbol u =0$) Navier-Stokes equation which in non-dimensional form reads:
\begin{align}
\label{eq:velo_eq}
    \mathrm{Re}_\mathrm{n}(\partial_t\boldsymbol u + \boldsymbol u \cdot\nabla\boldsymbol u) = &-\nabla p +\Delta\boldsymbol u - \mathrm{R}_\mathrm{f}\boldsymbol u\nonumber\\
    &+ \frac{1}{\mathrm{Er}}\nabla\cdot\left[ \boldsymbol \sigma_\mathrm{e} - \mathrm{R}_\mathrm{a}\boldsymbol\sigma_\mathrm{a}\right]\,\text{.}
\end{align}

In this non-dimensional form based on the nematic scales $m_\mathrm{n}$, $l_\mathrm{n}$ and $t_\mathrm{n}$ (cf.~Supplemental Material (SM)~\footnote{See Supplemental Material at [URL will be inserted by publisher] for details on the parameter regime and onset of active nematic turbulence, non-dimensionalization of the equations and the low-Reynolds-number approximation as well as definitions of length scales and energy and enstrophy budgets. The SM includes references~\cite{giomi2015geometry, hemingway2016correlation,thampi2013velocity,carenza2020cascade,tritton1998physical,Pope2000,davidson2015turbulence,giomi2012banding,urzay2017multi}}), the equation weights the inertial forces with the microscopic Reynolds number $\mathrm{Re}_\mathrm{n}=\rho u_\mathrm{n}l_\mathrm{n}/\eta$ and the elastic and active stresses with the inverse of the Ericksen number $\mathrm{Er}=\eta u_\mathrm{n}l_\mathrm{n}/K$, where $\rho$ is the mass density, $\eta$ the dynamic viscosity and $K$ an elastic constant. In contrast to classical fluids, active nematics feature an additional elastic stress $\boldsymbol{\sigma}_\mathrm{e}=-\lambda S \Delta \boldsymbol Q + (\Delta \boldsymbol Q)\cdot\boldsymbol{Q} - \boldsymbol{Q}\cdot(\Delta \boldsymbol Q) + \lambda S \boldsymbol Q(S^2-1)$~\footnote{In this work, we treat $\lambda S$ as varying in space and time (with $\lambda=\mathrm{const.}$, $S(\boldsymbol{x},t)$) as was done in~\cite{giomi2015geometry}. Since $\lambda S=\mathrm{const.}$ is used in other work~\cite{thampi2013velocity,hemingway2016correlation}, we tested the investigated statistics against both cases and found qualitatively identical results.} as well as the active stress $\boldsymbol{\sigma}_\mathrm{a} = \boldsymbol{Q}$, which couple the orientational field $\boldsymbol{Q}$ to the flow field. While the elastic stress describes the reaction of the flow to the particles' reorientation, the active stress models the impact of motility on the fluid flow~\cite{simha2002hydrodynamic}. The ratio $\mathrm{R_\mathrm{a}} = l^2_\mathrm{n}/l^2_\mathrm{a}$ of nematic and active length scales varies the relative strength between the stresses. It depends on the nematic $l^2_\mathrm{n}=K/C$ and the active length scale $l^2_\mathrm{a}=K/\alpha$ defined based on the material constants $K$ and $C$ and the activity $\alpha$. In addition to the original model~\cite{giomi2015geometry}, we add linear friction as a simple approximation to interactions between a two-dimensional active nematic layer and its surrounding. Surface friction has been found to influence quasi-two-dimensional active nematic layers experimentally~\cite{guillamat2016probing, guillamat2016control} as well as numerically~\cite{thampi2014active, thijssen2020active}. We control its influence with the non-dimensional friction number $\mathrm{R}_\mathrm{f}=l^2_\mathrm{n}/l^2_\mathrm{f}$, which we define via the friction length scale $l^2_\mathrm{f}=\eta/\mu$ based on the friction coefficient $\mu$.\\
The orientational order is described by the symmetric and traceless second-rank tensor $Q_{ij}=S(n_i\,n_j - \delta_{ij}/2)$, where $\boldsymbol{n}$ is a director and $S = \sqrt{2\mathrm{Tr}(\boldsymbol Q^2)}$ quantifies the local nematic order. The orientational field evolves in its non-dimensional form according to:
\begin{align}
\label{eq:alig_eq}
    \partial_t \boldsymbol{Q} + \boldsymbol{u}\cdot \nabla \boldsymbol{Q} = \lambda S \boldsymbol{E} - \boldsymbol{W}\boldsymbol{Q} + \boldsymbol{Q}\boldsymbol{W} + \Delta \boldsymbol Q - \boldsymbol Q(S^2-1)\,\text{,}
\end{align}
where $E_{ij}=(\partial_i\,u_j+\partial_j\,u_i)/2$ and $W_{ij}=(\partial_i\,u_j-\partial_j\,u_i)/2$ are the symmetric and antisymmetric parts of the velocity gradient, respectively. The alignment parameter $\lambda$ controls the particles' reaction to shear~\cite{hemingway2016correlation}. The system's relaxation to a uniformly aligned state is effectively described by the diffusion of boundaries via $\Delta \boldsymbol Q$ as well as local alignment via $\boldsymbol Q(S^2-1)$.

As a measure for the impact of inertial effects on the fluid flow, we focus on the self-advection term $\boldsymbol{u}\cdot\nabla\boldsymbol{u}$ in \eqref{eq:velo_eq} which has not been considered in previous work~\cite{giomi2011excitable,giomi2012banding,giomi2014defect,giomi2015geometry} due to the low-Reynolds-number approximation. For direct comparison, we performed numerical simulations on a periodic domain of size $L=N_x\Delta x$. We use a pseudo-spectral scheme for spatial discretization with a fourth-order Runge-Kutta scheme for integration in time. We scanned a range of values for the active number~$\mathrm{R}_\mathrm{a}$~which includes the onset of active nematic turbulence and ranges well into the turbulent regime (cf.~SM~\footnotemark[\value{footnote}], Tab.~S1). Within the studied parameter regime, we practically define the onset of active nematic turbulence by determining the lowest active number showing the creation and annihilation of defects ($\mathrm{R}_\mathrm{a}\approx0.05$, cf.~Fig.~\ref{fig:figure_3}a and SM~\footnotemark[\value{footnote}], Fig.~S1). As a starting point, we fix the microscopic Reynolds and the Ericksen number~$\mathrm{Re}_\mathrm{n}=\mathrm{Er}=0.1$~and vary the active number~$\mathrm{R}_\mathrm{a}$~over a wide range, similar to~\cite{giomi2015geometry}~(cf.~SM~\footnotemark[\value{footnote}], Tab.~S2). In order to study the statistics independent of the initial conditions, we equilibrated the system until it reached a statistically stationary state before analyzing the data. For $L=204.8$ we first ran simulations at a lower resolution $N_{x,\mathrm{eq}}=256$, $\Delta x_\mathrm{eq}=0.8$ to equilibrate the system for $N_{t,\mathrm{eq}}=5\cdot10^5$ time steps with $\Delta t_\mathrm{eq}=8\cdot10^{-3}$. We then upscaled and equilibrated the simulations to the final resolution $N_x= 512$, $\Delta x=0.4$ to resolve all spectral statistics well. We followed an analogous procedure to equilibrate the larger box size $L=819.2$ with $N_x=2048$. To achieve well-converged statistics, we averaged over an ensemble of $N_\mathrm{ens}=100$ realizations with independent random initial conditions as well as over the simulated time in the statistically stationary state.

\begin{figure}[!t]
    \centering
    \includegraphics[width=0.5\textwidth]{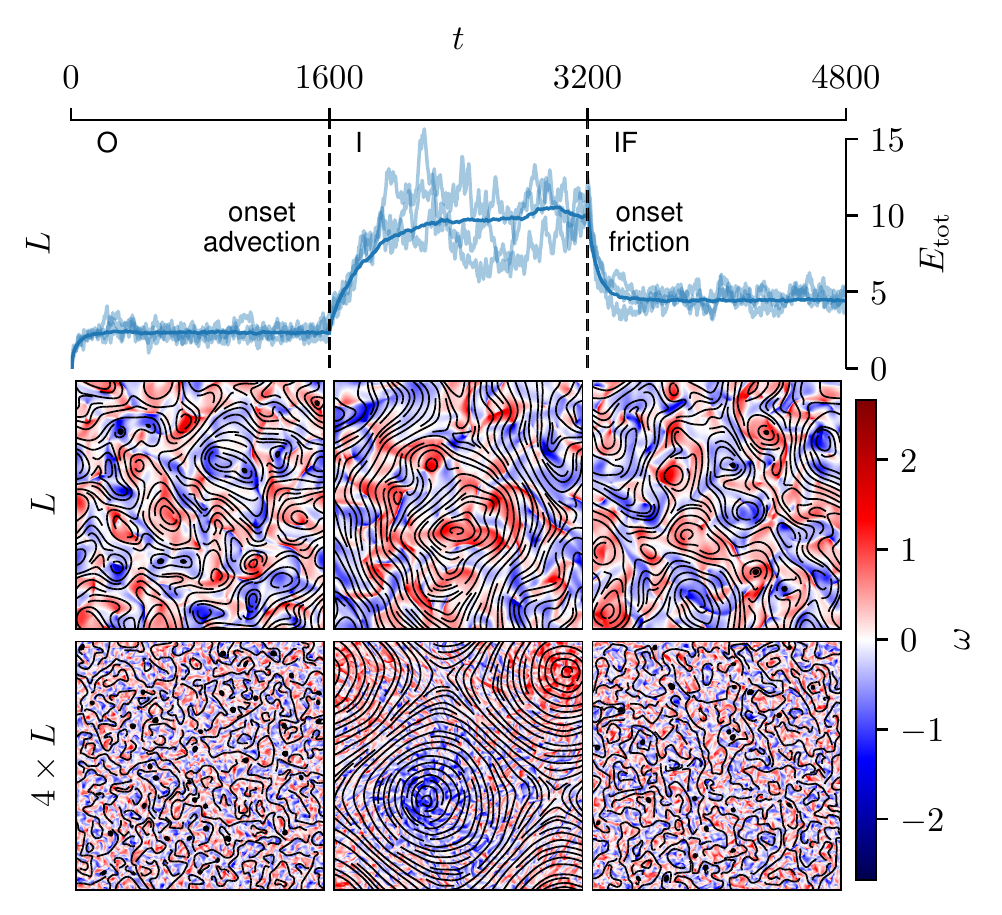}
    \caption{Advective inertia causes the formation of large-scale flow patterns and an increase in kinetic energy. Linear friction reduces these effects. Top: time evolution of the total kinetic energy $E_\mathrm{tot}$ in the original (O), inertial (I), and inertial with friction (IF) regime. Displayed are individual realizations (light blue) and the ensemble average (dark blue) with indicated times at which advection and friction are switched on (dashed lines). Center and bottom: snapshots of the vorticity $\omega=\partial_x u_y-\partial_y u_x$ (color map) and the velocity field (contour lines) taken from the statistically stationary states in each respective regime (box sizes $L$ and $4L$, $L=204.8$, $\mathrm{R}_\mathrm{a}=0.2$, see also Movies \href{https://youtu.be/clcnmBSx1c4}{S1}, \href{https://youtu.be/gcqqev76E1w}{S2}, and \href{https://youtu.be/3hy6agzCGWQ}{S3}).}
    \label{fig:figure_1}
\end{figure}

\begin{figure*}
    \centering
    \includegraphics[width=\textwidth]{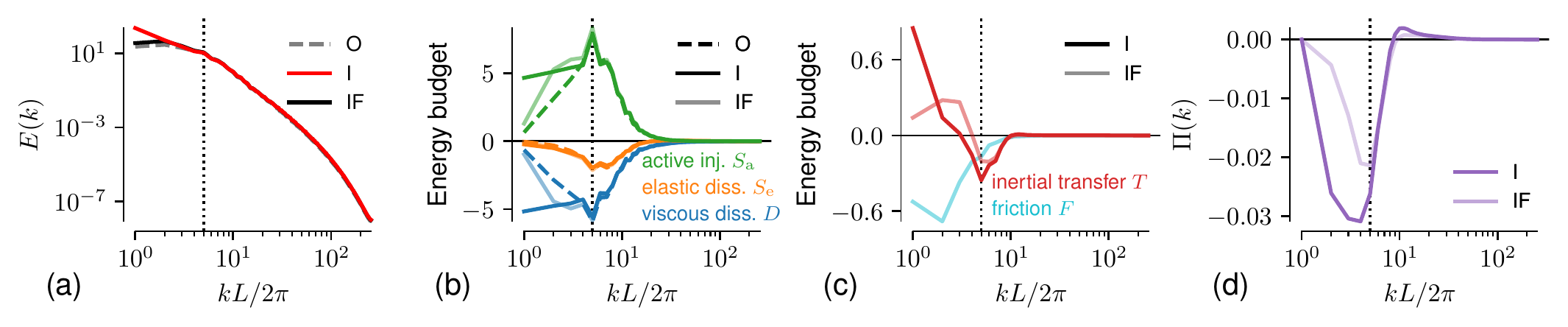}
    \caption{Advective inertia mediates an inverse energy transfer to the largest scales leading to an accumulation of energy and large-scale motion. Linear friction can compensate for this by directly dissipating the transferred energy. The original (O, dashed) regime is compared to the inertial (I, solid, dark) regime as well as to the regime with inertia and friction (IF, solid, light): ensemble- and time-averaged (a) kinetic energy spectrum $E(k)$, (b) and (c) energy budget contributions~\eqref{eq:def_energy_budget}, (d) energy flux $\Pi(k)$. The dotted lines indicate the scale with maximal active energy injection. ($\mathrm{R}_\mathrm{a}=0.2, L=204.8$)}
    \label{fig:figure_2}
\end{figure*}
Active stresses in this model for active nematic turbulence are able to induce inertial effects. A visualization of the main observations is given in Fig.~\ref{fig:figure_1}. For that, we computed an ensemble with a fixed set of parameters ($\mathrm{R}_\mathrm{a}=0.2$). This ensemble transitions over time from the original (O), over the inertial (I) regime, to the regime with inertia and friction (IF), by numerically switching on advection ($\boldsymbol{u}\cdot\nabla\boldsymbol{u}$) at the beginning of the first, and linear friction ($\mathrm{R}_\mathrm{f}>0$) at the beginning of the second transition. Changes from one to the next regime are clearly visible in the visualization as well as in the kinetic energy of the flow field.\\
Visually, the flow field in the original regime exhibits complex motion displayed by counter-rotating vortices (red and blue patches) which are interspersed by small vortex dipoles. The dipoles originate from the active stresses produced by topological defects in the orientational field and proliferate through the system. While these basic features persist in the inertial regime, the scale of motion increases drastically: rotating patterns on the scale of half the box size emerge. These patterns fluctuate strongly, decay, and reform over time (cf.~SM~\footnotemark[\value{footnote}], Movies \href{https://youtu.be/clcnmBSx1c4}{S1}, \href{https://youtu.be/gcqqev76E1w}{S2}, and \href{https://youtu.be/3hy6agzCGWQ}{S3}). If the simulation domain is large enough (compare results for $L$ and $4L$), they form meta-stable condensate-like patterns. The formation of homogeneous condensates, i.e.~large homogeneous vortices as known from classical two-dimensional turbulence~\cite{smith1993bose,smithr1994finite,boffetta2012two} and other continuum models of active turbulence~\cite{linkmann2019phase, linkmann2020condensate}, seems to be prevented by the small, proliferating vortex dipoles. However, the observations suggest that for a sufficiently large system, the disturbances by individual defects become small in comparison to the large-scale flow pattern, and a clearer and more stable condensate-like structure forms.\\
The kinetic energy in the original regime fluctuates around a mean value. At the onset of advection, however, the energy increases until it saturates to a higher mean value. While the initial increase in energy is accompanied by a build-up of more chaotic and longer-ranged motion, the new steady state corresponds to the fully developed flow with large-scale motion.

\begin{figure*}[!htb]
    \centering
    \includegraphics[width=\textwidth]{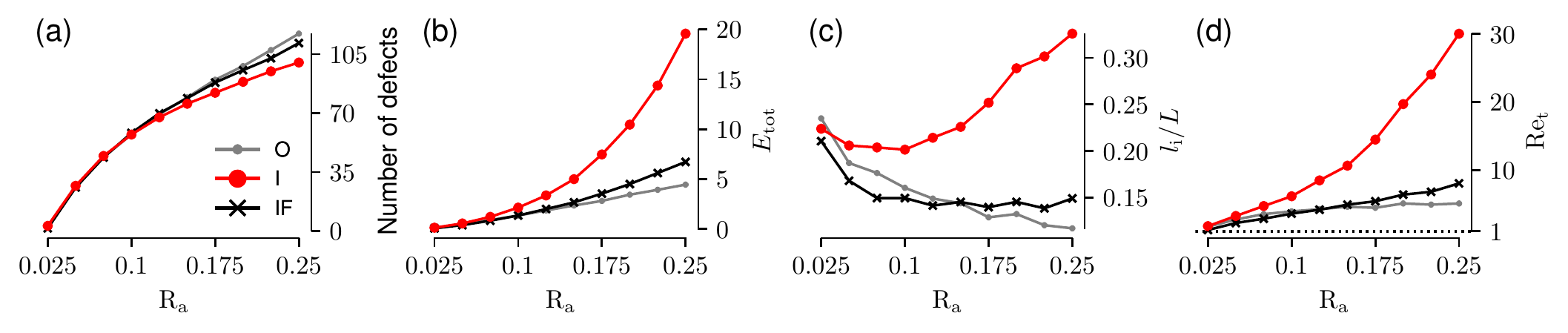}
    \caption{Advective inertia impacts statistics of the fluid flow already for weak and increasingly for stronger active forcing. The original (O) regime is compared to the inertial regime (I) as well as to the regime with inertia and friction (IF): ensemble- and time-averaged (a) number of defects, (b) total kinetic energy  $E_\mathrm{tot}$, (c) integral length $l_\mathrm{i}$, (d) turbulent Reynolds number $\mathrm{Re}_\mathrm{t}$ computed from (b) and (c). ($L=204.8$)}
    \label{fig:figure_3}
\end{figure*}
The appearance of large-scale motion together with the increase in kinetic energy demonstrate that the inclusion of advective inertia changes the dynamics and statistics of the fluid flow drastically. Furthermore, this suggests a connection between the scales of the flow and the increase in kinetic energy. 

Indeed, a spectral analysis of the kinetic energy, i.e.~its spectrum and budget, explains the observations (Fig.~\ref{fig:figure_2}).
The kinetic energy spectrum provides a scale-by-scale characterization of the kinetic energy and is defined as~\cite{alexakis2018cascades}:
\begin{align}
\label{eq:def_energy_spectrum}
    E(k, t) = \frac{1}{2\Delta k} \sum_{k\le|\boldsymbol{k}|<k+\Delta k} |\hat{\boldsymbol{u}}(\boldsymbol{k}, t)|^2\,\text{,}
\end{align}
where $\Delta k=2\pi/L$. Comparing the original and inertial regime (Fig.~\ref{fig:figure_2}a), the spectrum displays a prominent rise in energy at small wave numbers, i.e.~large scales. In contrast, energy increases only a little at higher wave numbers corresponding to smaller scales. Consequently, the increase in total kinetic energy in the inertial regime primarily stems from energy accumulating at large scales. This fits well with the observation of dominant large-scale motion.\\
The spectral energy budget allows to study scale by scale how much energy each term in \eqref{eq:velo_eq} injects into or dissipates from the flow:
\begin{align}
\label{eq:def_energy_budget}
    \partial_t\,E = T + D + S_\mathrm{e} + S_\mathrm{a} + F\,\text{,}
\end{align}
where $T(k, t)$ is the inertial energy transfer due to advection, $D(k, t)$ the viscous dissipation, $S_\mathrm{e}(k, t)$ the elastic dissipation, $S_\mathrm{a}(k, t)$ the active injection, and $F(k, t)$ the dissipation through friction (cf.~SM~\footnotemark[\value{footnote}], Sec.~S3). Interestingly, energy injection by active stresses occurs on a broad range of scales and features a maximum (Fig.~\ref{fig:figure_2}b). In the steady state of the original regime, viscous and elastic forces dissipate the injected energy, i.e.~the three contributions balance each other at each scale~\footnote{In addition to active injection, the elastic stress also injects energy at small scales, which becomes more visible in the enstrophy budget. (cf.~SM, Fig.~S2 or~\cite{carenza2020cascade})}. In the inertial regime however, the advection term mediates an energy transfer between scales (Fig.~\ref{fig:figure_2}c). It extracts energy around the scale of maximal injection and transfers the majority towards larger and a small portion towards smaller scales. This is quantified by the energy flux (Fig.~\ref{fig:figure_2}d):
\begin{align}
    \Pi(k, t) = \int^{\infty}_{k} \mathrm{d}{k'} \, T(k', t)\,\text{,}
\end{align}
where a negative flux indicates an inverse transfer to larger and a positive flux a direct transfer to smaller scales. This flux of energy between scales leads to an overall increase in energy, which is compensated for on small scales by viscous dissipation but remains initially unbalanced on large scales. Only after an initial accumulation of energy on large scales, viscous dissipation becomes strong enough to compensate for any further flux mediated by advective inertia. Interestingly, the active stress contributions result in an even larger injection on these scales when the large-scale energy is increased, which is balanced by a similar increase in viscous dissipation (Fig.~\ref{fig:figure_2}b). Remarkably, the peak magnitude of the energy transfer due to advection is more than one order of magnitude smaller than the peak magnitudes of active stress contributions and viscous dissipation, indicating that inertial advection has a comparably small effect on the flow at any particular instant, which however builds up over time.

The spectral analysis demonstrates that advective inertia is responsible for an inverse transfer of energy towards larger scales, at which energy accumulates, resulting in the observed large-scale motion. To quantify how the strength of active forcing mediates the inertial effects, we vary $\mathrm{R}_\mathrm{a}$ ranging from the onset of active nematic turbulence until deep into the turbulent regime.\\
Recall that we define the onset of active nematic turbulence, where defects first spontaneously form and annihilate ($\mathrm{R}_\mathrm{a}\approx0.05$, cf.~Fig.~\ref{fig:figure_3}a and SM~\footnotemark[2], Fig.~S1). The number of defects~\footnote{The defect detection algorithm described in~\cite{huterer2005distribution} was used.} increases with activity in both regimes as can be expected for increased disorder due to stronger flow. Interestingly, it is smaller in the inertial than in the original regime.\\
As indicated in Fig.~\ref{fig:figure_1}, the total kinetic energy~$E_\mathrm{tot}(t)=\sum_k E(k, t)\,\Delta k$~is higher in the inertial than in the original regime. As can be expected for stronger active driving, the difference grows larger with activity (Fig.~\ref{fig:figure_3}b). Interestingly, it is already nonzero at the onset of active nematic turbulence and grows continuously. This suggests that inertial effects are non-negligible for all activities studied in this parameter regime.\\
The turbulent Reynolds number based on the integral scale is a typical measure for the importance of inertial effects in comparison to viscous dissipation in turbulent flows:~$\mathrm{Re}_\mathrm{t} = \sqrt{E_\mathrm{tot}}l_\mathrm{i}/\nu$~, where $\nu = \eta/\rho$ is the kinematic viscosity and $l_\mathrm{i}$ is the integral length scale (cf.~SM~\footnotemark[2]). Indeed, in both regimes, the emerging Reynolds number is larger than unity for all activities (Fig.~\ref{fig:figure_3}d), which means that inertial effects are not negligible (cf.~SM~\footnotemark[2]). The difference in Reynolds numbers between the original and the inertial regime increases, as the kinetic energy, with activity. This is readily understood because the turbulent Reynolds number is proportional to the root of the kinetic energy, and because the integral length scale increases with the emergence of large-scale patterns in the inertial regime (Fig.~\ref{fig:figure_3}c).

The parameter scan for the activity demonstrates that advective inertia changes the fluid flow increasingly with activity, already starting at the onset of active nematic turbulence. Consistent modeling, therefore, requires the inclusion of the advection term in this parameter regime. This together with the absence of large-scale motion in form of condensate-like flow patterns in experiments motivates the addition of linear friction to the model (regime with inertia and friction in Fig.~1). Linear friction dissipates energy primarily on large scales, which contain the most energy (Fig.~\ref{fig:figure_2}c). It thereby counteracts the accumulation of energy through the inverse transfer and prohibits the formation of large-scale flow (Fig.~\ref{fig:figure_1} and Fig.~\ref{fig:figure_2}a). While it can restore the system to a state very similar to the original regime (Fig.~\ref{fig:figure_3}), i.e. without advection or friction, its impact on the flow depends on the ratio of friction coefficient and activity.

\begin{figure}[!ht]
    \centering
    \includegraphics[width=0.5\textwidth]{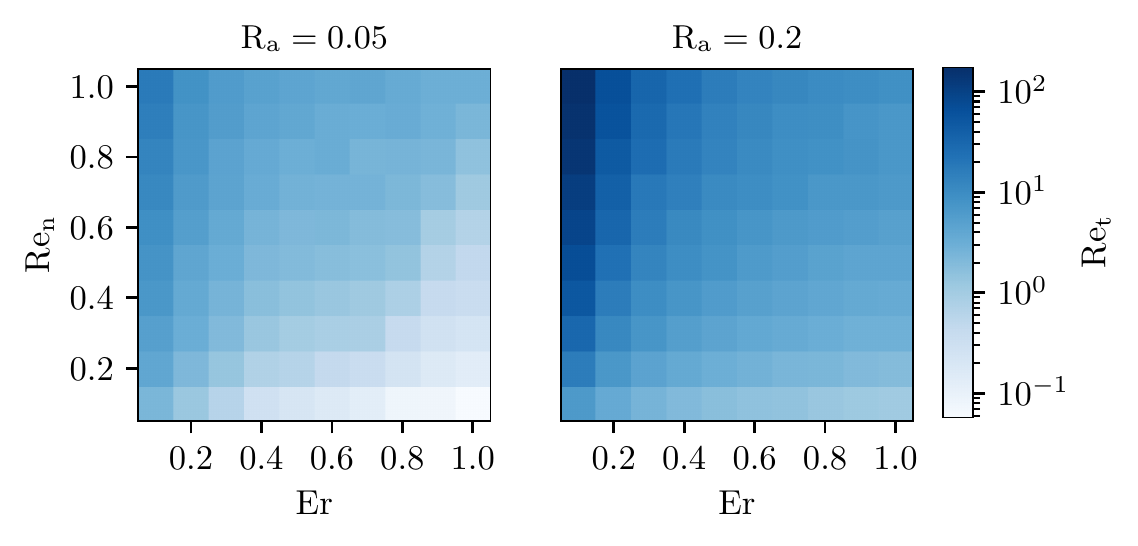}
    \caption{The importance of inertia, quantified by the turbulent Reynolds numbers $\mathrm{Re}_\mathrm{t}$, increases directly with the microscopic Reynolds number $\mathrm{Re}_\mathrm{n}$ and inversely with the Ericksen number. The simulations were performed in the regime with inertia and friction and were averaged over $30$ independent random initial conditions. ($\mathrm{R}_\mathrm{f}=7.5\cdot10^{-4}, L=204.8$)}
    \label{fig:figure_4}
\end{figure}

So far, we have varied the active forcing strength at fixed microscopic Reynolds and Ericksen numbers. To get a broader overview, Fig.~\ref{fig:figure_4} shows the turbulent Reynolds number as a function of the microscopic Reynolds number and the Ericksen number for two different activities in the case including inertia and friction. For small microscopic Reynolds numbers and comparably large Ericksen numbers, the turbulent Reynolds number is generally small, consistent with recent findings~\cite{carenza2020cascade}~(cf.~SM~\footnotemark[\value{footnote}], Tab.~S2, Fig.~S3).
However, as the Ericksen number is decreased, the turbulent Reynolds number increases. The largest turbulent Reynolds numbers emerge for the smallest Ericksen numbers, where they can have values significantly larger than one even if the microscopic Reynolds number is small. In this regime, active stresses can effectively drive the flow, thereby exciting inertial effects.

To summarize, by comparing a continuum model for active nematic turbulence with and without inertial advection, we find that inertial effects can significantly alter the fluid flow by an accumulation of kinetic energy on the largest scales of the flow. In particular, we observe large-scale motion in form of condensate-like flow patterns when inertial effects are present. Remarkably, inertial effects start to play a role already for small activities, provided the active stresses can effectively drive the flow.

Linear friction, included to model surface friction present in the experimental setups~\cite{guillamat2016probing}, compensates for the effects of inertial advection by dissipating the inversely transferred energy, resulting effectively in a flow similar to the system without advection or friction. Including inertial advection and friction appears therefore necessary for a consistent theoretical description of active nematic turbulence in certain parameter ranges.

\section{Acknowledgments}
We thank the reviewers for helpful comments. This work was supported by the Max Planck Society. M.W. gratefully acknowledges a Fulbright-Cottrell Award grant.

\bibliographystyle{apsrev4-1}
\bibliography{literature}

\clearpage
\onecolumngrid
\setcounter{page}{1}
{
\centering
\date{\today}
    {\large \textbf{Supplementary Material for \\``The role of advective inertia in active nematic turbulence''}}

    \vspace{.5cm}
    Koch et al.
    
    \vspace{.5cm}
    (Dated: \today)
    
}

\counterwithout*{equation}{section}
\renewcommand{\figurename}{Supplementary Figure}
\renewcommand{\tablename}{Supplementary Table}
\renewcommand{\thesection}{\arabic{section}}
\setcounter{table}{0}
\renewcommand{\thetable}{S\arabic{table}}%
\setcounter{figure}{0}
\renewcommand{\thefigure}{S\arabic{figure}}%
\setcounter{equation}{0}
\def\theequation{S\arabic{equation}}
\setcounter{section}{0}
\titleformat{\section}{\centering\small\bfseries\uppercase}{Supplementary Note \thesection.}{1em}{}

\section{Non-dimensionalization of the equations of motion} \label{app:nondim}
For comparison of parameter choices investigated in this paper and in previous literature, we obtain in the following the set of non-dimensional equations stated in the main text by rescaling length, time and mass with nematic scales $l_\mathrm{n}$, $t_\mathrm{n}$, and $m_\mathrm{n}$, respectively. We first state the dimensional equations of motion, then motivate the nematic scales and rescale the equations with them. We finally identify the dimensionless numbers and discuss the meaning of this particular choice of non-dimensionalization.

\subsection{Dimensional equations of motion}
\label{app:dim_eq_of_motion}
The Navier-Stokes equation we use to model a 2D active nematic (as adapted from~\cite{giomi2015geometry}) is:
\begin{align} \label{eq:SM_dim_NS}
    \nabla \cdot \boldsymbol u &= 0 \nonumber \\
    \rho(\partial_t\boldsymbol u + \boldsymbol u \cdot \nabla \boldsymbol u) &= -\nabla p + \eta \Delta \boldsymbol u + \nabla\cdot (\boldsymbol \sigma_\mathrm{e} - \boldsymbol \sigma_\mathrm{a}) - \mu \boldsymbol u\,\text{,}\\
    \boldsymbol \sigma_\mathrm{e} &= -\lambda S \boldsymbol H + \boldsymbol H \boldsymbol Q - \boldsymbol Q \boldsymbol H\,\text{,}\nonumber\\
    \boldsymbol \sigma_\mathrm{a} &= \alpha\boldsymbol Q\,\text{,}\nonumber
\end{align}
where $\boldsymbol \sigma_\mathrm{e}$ and $\boldsymbol \sigma_\mathrm{a}$ are the elastic and active stresses, respectively, and $\boldsymbol H$ is the molecular tensor defined as the functional derivative of the free energy:
\begin{align}
\label{eq:mole_tens}
    \boldsymbol H &= K\Delta\boldsymbol Q - C\boldsymbol Q (S^2 -1) = -\frac{\delta \mathcal{F}}{\delta\boldsymbol Q}\,\text{,}\\
    \mathcal{F} &= \int \mathrm{d}^2r\,\frac{K}{2} (\partial_i\,Q_{jk})^2 + \frac{C}{2}((Q_{ij}\,Q_{ji})^2 - Q_{ij}\,Q_{ji})\,\text{,}\nonumber\\
    S &= \sqrt{2\,Q_{ij}\,Q_{ji}}\,\text{.}\nonumber
\end{align}
The time evolution for the alignment tensor is given by:
\begin{align}
\label{eq:alig_tens_dim}
    \partial_t \boldsymbol Q + \boldsymbol u\cdot\nabla\boldsymbol Q &= \lambda S \boldsymbol E - \boldsymbol W\boldsymbol Q + \boldsymbol Q\boldsymbol W + \gamma^{-1}\boldsymbol H\,\text{,}\\
    E_{ij} &= \frac{1}{2}(\partial_i u_j + \partial_j u_i)\,\text{,}\nonumber\\
    W_{ij} &= \frac{1}{2}(\partial_i u_j - \partial_j u_i)\,\text{.}\nonumber
\end{align}
The definition of the parameters can be found in Tab.~\ref{tab:para}.

\subsection{Nematic scales}
For our non-dimensionalization we use the nematic length, time and mass scales
\begin{align}
\label{eq:SM_nematic_scales}
    l_\mathrm{n} = \sqrt{\frac{K}{C}}\,\text{,}\qquad t_\mathrm{n} = \frac{\gamma}{C}\,\text{,}\quad\text{and}\quad m_\mathrm{n} = \frac{\gamma^2}{C}\,\text{,}
\end{align}
which are defined with respect to parameters controlling the relaxation towards a uniformly aligned state. Namely, these are the elastic constant~$K$, the material constant~$C$~and the rotational viscosity $\gamma$, which control the orientational diffusion, alignment, and damping of~$\boldsymbol Q$~in eq.~\eqref{eq:mole_tens}~and eq.~\eqref{eq:alig_tens_dim}, respectively. \\
The nematic length scale~$l_\mathrm{n}$~can be understood to be proportional to the defect core radius when the system is deep in the nematic regime~\cite{hemingway2016correlation}. From observing the vorticity and order parameter fields, defects seem to be the smallest visible structures in the active nematic turbulent state (Fig.~\ref{fig:figure_1_app}~and Fig.~1~in the main text). To resolve their scales properly, it is reasonable to use the nematic length scale for measuring length scales in the system.\\
The nematic time scale~$t_\mathrm{n}=\gamma\, l^2_\mathrm{n}/K$~can be thought of as the time scale over which distortions in the orientational field relax to uniform alignment on length scales~$l_\mathrm{n}$~\cite{hemingway2016correlation}. It may therefore be accessible experimentally as a characteristic scale in the system without activity.\\
For the nematic mass, motivation from observations is more difficult. However, choosing the same parameters appearing in the nematic length and time scales, our definition of the nematic mass is consistent and unique.

\subsection{Rescaling the equations of motion}
In the following, we rescale all fields with the nematic length, time and mass scales defined in the last section. For conciseness, we additionally use velocities~$u_\mathrm{n} = l_\mathrm{n} / t_\mathrm{n}$. The non-dimensional variables are denoted with an asterisk (in the main text and the following sections, the asterisk is dropped).\\
The velocity, as well as spatial and temporal derivatives, are:
\begin{align}
    \boldsymbol{u} = u_\mathrm{n}\,\boldsymbol{u}^*\,\text{, }\, \nabla = \frac{1}{l_\mathrm{n}}\nabla^*\,\text{, }\, \Delta = \frac{1}{l^2_\mathrm{n}}\Delta^*\,\text{, }\, \partial_t=\frac{1}{t_\mathrm{n}}\partial_{t^*}\,\text{.}
\end{align}
To rescale the pressure one typically chooses one of the following two options~\cite[p.~92]{tritton1998physical}:
\begin{align}
    \frac{p}{\rho} \equiv \Tilde{p} = u^2_\mathrm{n}\,\Tilde{p}^*\,\qquad\text{or}\,\qquad \frac{p}{\eta} \equiv \Tilde{p} = \frac{1}{t_\mathrm{n}}\,\Tilde{p}^*\,\text{,}
\end{align}
whereby the first option is used for flows in which inertia dominates and the latter for flows in which viscous dissipation dominates. Because the pressure gradient accounts for the incompressibility of the flow, both choices aim at balancing the larger term against the pressure gradient. Here, expecting experimentally a dominance of viscous dissipation, we use the option in which pressure is rescaled with viscosity.\\
We split the elastic stress for clarity in two non-dimensional contributions, which are subscripted with their prefactors $K$ and $C$. In addition, we identify the non-dimensional active stress as the alignment tensor with the activity parameter as prefactor:
\begin{align}
    \boldsymbol{\sigma}_\mathrm{e} &= \frac{K}{l^2_\mathrm{n}} \left[-\lambda S\Delta^*\boldsymbol{Q} + (\Delta^*\boldsymbol{Q})\boldsymbol{Q} - \boldsymbol{Q}(\Delta^*\boldsymbol{Q})\right] + C\lambda S\boldsymbol{Q}(S^2-1) = \frac{K}{l^2_\mathrm{n}}\boldsymbol\sigma^*_K + C\boldsymbol\sigma^*_C\,\text{,}\\
    \boldsymbol{\sigma}_\mathrm{a} &= \alpha \boldsymbol{Q} = \alpha\boldsymbol{\sigma}^*_\mathrm{a} \,\text{.}
\end{align}
The non-dimensionalized evolution equations~\eqref{eq:SM_dim_NS}~and~\eqref{eq:alig_tens_dim}~then read:
\begin{align}
    \rho\left(\frac{u_\mathrm{n}}{t_\mathrm{n}}\,\partial_{t^*}\boldsymbol{u}^* + \frac{u^2_\mathrm{n}}{l_\mathrm{n}}\,\boldsymbol{u}^*\cdot \nabla^*\boldsymbol{u}^*\right) &= -\frac{\eta}{t_\mathrm{n}\,l_\mathrm{n}}\,\nabla^*\Tilde{p}^* + \frac{\eta\,u_\mathrm{n}}{l^2_\mathrm{n}}\,\Delta^*\boldsymbol{u}^* + \frac{1}{l_\mathrm{n}}\nabla^*\cdot\left[ \frac{K}{l^2_\mathrm{n}}\boldsymbol{\sigma}^*_K + C\boldsymbol{\sigma}^*_C - \alpha\boldsymbol{\sigma}^*_\mathrm{a}\right] - \mu \,u_\mathrm{n}\,\boldsymbol{u}^*\,\text{,}\\
    \frac{1}{t_\mathrm{n}}\,\partial_{t^*} \boldsymbol{Q} + \frac{u_\mathrm{n}}{l_\mathrm{n}} \boldsymbol{u}^*\cdot\nabla^*\boldsymbol{Q} &= \frac{u_\mathrm{n}}{l_\mathrm{n}}\lambda S \boldsymbol{E}^* - \frac{u_\mathrm{n}}{l_\mathrm{n}}\boldsymbol{W}^* \boldsymbol{Q} + \frac{u_\mathrm{n}}{l_\mathrm{n}}\boldsymbol{Q}\boldsymbol{W}^* + \frac{K}{\gamma\,l^2_\mathrm{n}}\Delta^*\boldsymbol{Q} - \frac{C}{\gamma}\boldsymbol{Q} (S^2 - 1)\,\text{.}
\end{align}
Until here, the specific choice of length, time, and mass scales did not matter. However, now we can simplify the expressions by choosing specifically the nematic scales~\eqref{eq:SM_nematic_scales} and obtain:
\begin{align}
\label{eq:nondimensional_navier_stokes}
    \frac{\rho\,u_\mathrm{n}\,l_\mathrm{n}}{\eta}(\partial_{t^*}\boldsymbol{u}^* + \boldsymbol{u}^*\cdot \nabla^*\boldsymbol{u}^*) &= -\nabla^*\Tilde{p}^* + \Delta^*\boldsymbol{u}^* + \frac{K}{\eta\,u_\mathrm{n}\,l_\mathrm{n}}\nabla^*\cdot\left[ \boldsymbol{\sigma}^*_\mathrm{e} - \frac{\alpha\,l^2_\mathrm{n}}{K}\boldsymbol{\sigma}^*_\mathrm{a}\right] - \frac{\mu\,l^2_\mathrm{n}}{\eta}\boldsymbol{u}^*\,\text{,}\\
    \partial_{t^*} \boldsymbol{Q} + \boldsymbol{u}^*\cdot\nabla^*\boldsymbol{Q} &= \lambda S \boldsymbol{E}^* - \boldsymbol{W}^* \boldsymbol{Q} + \boldsymbol{Q}\boldsymbol{W}^* + \Delta^*\boldsymbol{Q} - \boldsymbol{Q} (S^2 - 1)\,\text{,}
\end{align}
where we write now the non-dimensional elastic stress $\boldsymbol\sigma^*_\mathrm{e}=\boldsymbol\sigma^*_K + \boldsymbol\sigma^*_C$. This equation leaves us with four dimensionless prefactors (plus the flow alignment parameter $\lambda$) which we define as the four dimensionless numbers
\begin{align}
    \mathrm{Re}_\mathrm{n} = \frac{\rho\,u_\mathrm{n}\,l_\mathrm{n}}{\eta}\,\text{,}\qquad \mathrm{Er} = \frac{\eta\,u_\mathrm{n}\,l_\mathrm{n}}{K}\,\text{,}\qquad \mathrm{R}_\mathrm{a} = \frac{l^2_\mathrm{n}}{l^2_\mathrm{a}}\,\text{,}\quad\text{and}\quad \mathrm{R}_\mathrm{f} = \frac{l^2_\mathrm{n}}{l^2_\mathrm{f}}\,\text{.}
\end{align}
We call $\mathrm{Re}_\mathrm{n}$ the microscopic (nematic) Reynolds number, $\mathrm{Er}$ the Ericksen number, $\mathrm{R}_\mathrm{a}$ the active and $\mathrm{R}_\mathrm{f}$ the friction number. The latter two are ratios of length scales:
\begin{align}
    l_\mathrm{n} = \sqrt{\frac{K}{C}}\,\text{,}\qquad l_\mathrm{a} = \sqrt{\frac{K}{\alpha}} \,\text{,}\qquad l_\mathrm{f} = \sqrt{\frac{\eta}{\mu}}\,\text{,}
\end{align}
where $l_\mathrm{n}$ is the nematic, $l_\mathrm{a}$ is the active and $l_\mathrm{f}$ is the friction length scale~\cite{giomi2015geometry}.\\
Thus, by choosing the nematic scales to non-dimensionalize the equations of motion, we effectively reduced the number of free parameters from seven to four (not counting the flow alignment parameter~$\lambda$).\\
Note at this point that the above-defined Reynolds and Ericksen numbers are defined via the nematic length and time scales. Different definitions of these numbers are possible, which means that the values of these numbers can be interpreted only in relation to the scales used to define them (see discussion below).

\subsection{Discussion of the non-dimensionalization}
In this study, we choose to non-dimensionalize the equations of motion with the nematic scales. Different scales can be used which will result in different prefactors than the ones identified in the last section. A particular useful choice is based on scales that are representative of the actual characteristic scales in the problem, which allows predicting whether one term in the equations of motion dominates over another. We will call this particular choice of scales the characteristic scales in the following. Using the characteristic scales, each term in the equations of motion is normalized, i.e.~it only takes values of order one. For example, the nonlinear advection~$\boldsymbol u \cdot \nabla \boldsymbol u$ without its prefactor $\mathrm{Re}_\mathrm{n}$ as well as the viscous diffusion~$\Delta\boldsymbol u$ would be of order one. The microscopic Reynolds number~$\mathrm{Re}_\mathrm{n}$ then characterizes how strong nonlinear advection contributes to the time evolution compared to viscous diffusion. Depending on the actual value of the characteristic scales, the Reynolds number may be very small. In this case, nonlinear advection would contribute only little to the dynamics and could be neglected, justifying a low-Reynolds number approximation. The nematic scales used in the non-dimensionalization shown above are not characteristic scales and do not necessarily normalize each term in the equations of motion. Hence, no a priori information can be used to simplify the equations of motion.

In 2D active nematic turbulence, the characteristic scales of the flow and the orientational field are a priori unknown. While individual constituents, such as microtubule proteins, may define a specific microscopic length scale, the flow and orientational fields are excited on much larger scales, presumably better defined by the topological defects in the system. It is, therefore, questionable to apply the low-Reynolds approximation without the knowledge of the characteristic scales. The result can be, as has been shown in this paper, that parameter ranges are chosen which excite the flow strongly enough that the low-Reynolds number approximation becomes invalid.

Note that the microscopic Reynolds number, which we identified above as the prefactor of the inertial terms, is not the only Reynolds number that can be defined. In the main text, for instance, we compute the turbulent Reynolds number~$\mathrm{Re}_\mathrm{t} = \sqrt{E_\mathrm{tot}}\,l_\mathrm{i}/\nu$~, where $\nu = \eta/\rho$ is the kinematic viscosity and $l_\mathrm{i}$ is the integral length scale defined via the longitudinal velocity correlation function~\cite[p.~197]{Pope2000}: 
\begin{align}
    l_\mathrm{i} = \int^{\infty}_0 \mathrm{d}r\,\frac{\langle u_x(\boldsymbol{x}+r\boldsymbol{e}_x, t)\,u_x(\boldsymbol{x}, t)\rangle}{\langle u_x(\boldsymbol{0}, t)^2\rangle}\,\text{,}
\end{align}
which characterizes the flow field based on the emerging velocities and the scale over which velocities are correlated.

 \section{Parameters and onset of active nematic turbulence}
\label{app:parameter}
\begin{figure}
    \centering
    \includegraphics[width=0.6\textwidth]{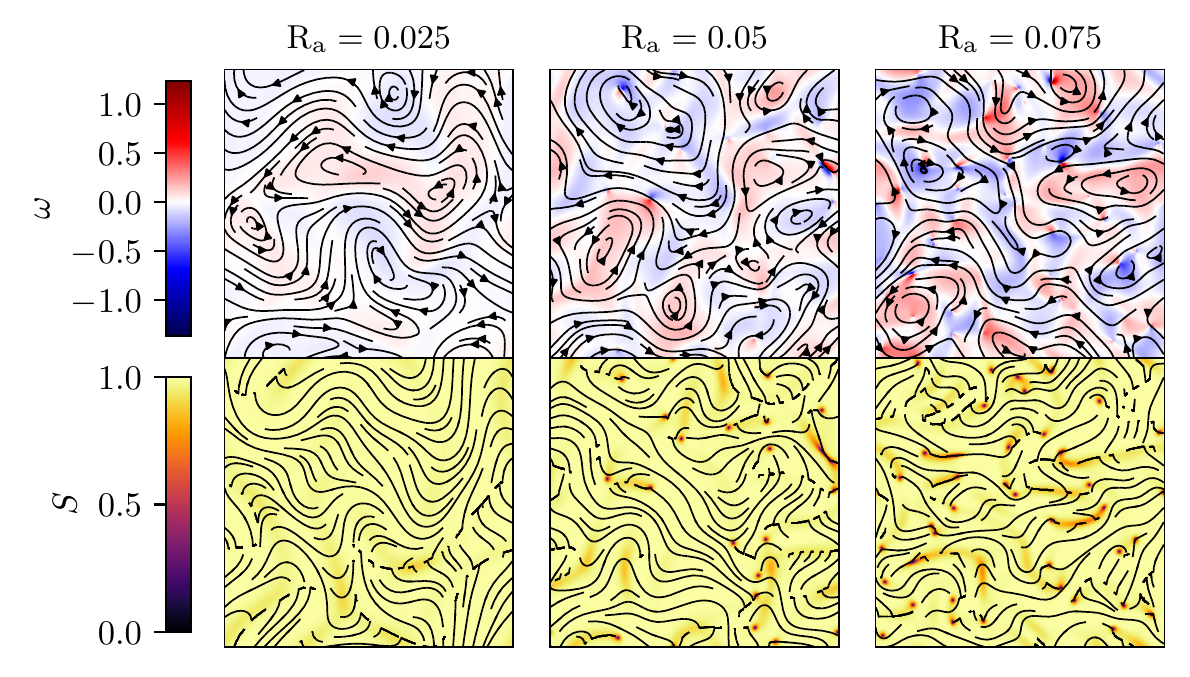}
    \caption{Active stresses excite flows that strongly bend the orientational field. For sufficiently large activities, pairs of defects form (corresponding to zero order parameter), move through the system, and annihilate. We identify the onset of active turbulence close to $\mathrm{R}_\mathrm{a}=0.05$, since this is (by magnitude) the smallest investigated activity parameter for which we observe the emergence of defects. Top: vorticity (color map) and velocity (streamlines) for activities close to the onset of active nematic turbulence. Bottom: local nematic order parameter (color map) and director (streamlines). ($\mathrm{R}_\mathrm{f}=0$, $L=204.8$)}
    \label{fig:figure_1_app}
\end{figure}

The parameters used in this paper are listed in Tab.~\ref{tab:para}. As with any numerical simulation, the physical dimensions, i.e.~units, need to be specified such that a comparison to a real physical system becomes possible. Here, we specify all parameters in terms of the nematic scales defined above, i.e. the nematic length $l_\mathrm{n}$, time $t_\mathrm{n}$, mass $m_\mathrm{n}$, and velocity $u_\mathrm{n} = l_\mathrm{n} / t_\mathrm{n}$. The values listed in Tab.~\ref{tab:para} are the rescaled ones, where the numerical values in simulation units can be obtained by using $l_\mathrm{n}=0.05, t_\mathrm{n}=0.025, m_\mathrm{n}=0.25$. Since numerical values for continuum-model parameters are not consistent throughout studies of active nematic turbulence, we compare our parameter values to a selection of other papers (Tab.~\ref{tab:para_comparison}).\\
We define the onset of active nematic turbulence within our parameter regime as the value of activity for which topological defects are first created (Fig.~\ref{fig:figure_1_app}).

\begin{table}
    \centering
    \ra{1.2}
    \begin{tabular}{@{}p{0.1\textwidth}p{0.25\textwidth}p{0.25\textwidth}p{0.15\textwidth}@{}}
    \hline
        Parameter & Description & Numerical Value & Dimensions\\
    \hline
        $\rho$ & Solvent density & $0.01$ & $\left[m_\mathrm{n}\right]\left[l_\mathrm{n}\right]^{-2}$\\
        & & Fig.~4, S3: $0.01\xrightarrow{}1$ & \\
        $\eta$ & Solvent viscosity & $0.1$ & $\left[m_\mathrm{n}\right]\left[t_\mathrm{n}\right]^{-1}$\\
        & & Fig.~4, S3: $0.01\xrightarrow{}1$ & \\
        $K$ & Elastic constant & $1.0$ & $\left[m_\mathrm{n}\right]\left[u_\mathrm{n}\right]^{2}$\\
        $C$ & Energy density scale & $1.0$ & $\left[m_\mathrm{n}\right]\left[t_\mathrm{n}\right]^{-2}$\\
        $\gamma$ & Rotational viscosity & $1.0$ & $\left[m_\mathrm{n}\right]\left[t_\mathrm{n}\right]^{-1}$\\
        $\lambda$ & Alignment parameter & $0.1$ & $-$\\
        $\alpha$ & Activity & $0.025\to0.25$ & $\left[m_\mathrm{n}\right]\left[t_\mathrm{n}\right]^{-2}$\\
        $\mu$ & Linear friction coefficient & $0.0, 7.5\cdot10^{-5}$ & $\left[m_\mathrm{n}\right]\left[l_\mathrm{n}\right]^{-2}\left[t_\mathrm{n}\right]^{-1}$\\
        & & Fig.~4, S3: $7.5\cdot10^{-5}\xrightarrow{}7.5\cdot10^{-4}$ & \\
        $\Delta x$ & Grid spacing & $0.4$ & $\left[l_\mathrm{n}\right]$\\
        $\Delta t$ & Time step width & $0.002$ & $\left[t_\mathrm{n}\right]$\\
        $N_x$ & Number of grid points & $512, 2048$ & $-$\\
        $L$ & Box size & $204.8, 819.2$ & $\left[l_\mathrm{n}\right]$\\
    \hline
    \end{tabular}
    \caption{Model and simulation parameters as well as their values. Dimensions are defined via the nematic mass $m_\mathrm{n}$, length $l_\mathrm{n}$ and time $t_\mathrm{n}$. These parameters were used in the numerical code and are equivalent to the non-dimensional numbers shown in Tab.~\ref{tab:para_comparison}.}
    \label{tab:para}
\end{table}{}

\begin{table*}
    \centering
    \ra{1.2}
    \begin{tabular}{@{}p{0.2\textwidth}p{0.05\textwidth}p{0.05\textwidth}p{0.07\textwidth}p{0.0225\textwidth}p{0.07\textwidth}p{0.1\textwidth}p{0.07\textwidth}p{0.05\textwidth}p{0.07\textwidth}@{}}
    \hline
         & $\mathrm{Re}_\mathrm{n}$ & $\mathrm{Er}$ & \multicolumn{3}{l}{$\mathrm{R}_\mathrm{a}$} & $\mathrm{R}_\mathrm{f}$ & $L$ & $N_x$ & $\Delta x$\\
    \hline
        this paper & $0.1$ & $0.1$ & $0.025$ & $\to$ & $0.25$ & $7.5\cdot10^{-5}$ & $204.8$ & $512$ & $0.4$\\
        Giomi~\cite{giomi2015geometry} & $0.1$ & $0.1$ & $0.05$ & $\to$ & $2.5$ & $-$ & $200$ & $256$ & $0.78$\\
        Hemingway et al.~\cite{hemingway2016correlation} M1 & $0.1$ & $0.1$ & $10$ & $\to$ & $10^3$ & $-$ & $128$ & $128$ & $1.0$\\
        Hemingway et al.~\cite{hemingway2016correlation} M2 & $0$ & $1.76$ & $0.05$ & $\to$ & $12.8$ & $-$ & $707$ & $2048$ & $0.35$\\
        Thampi et al.~\cite{thampi2013velocity} & $0.01$ & $0.23$ & $0.003$ & $\to$ & $0.083$ & $-$ & $1095$ & $400$ & $2.74$\\
        Carenza et al.~\cite{carenza2020cascade} & $0.05$ & $0.83$ & $0.000625$ & $\to$ & $0.0625$ & $-$ & $2290$ & $512$ & $4.47$\\
        Urzay et al.~\cite{urzay2017multi} & $0.21$ & $0.22$ & $0.12$ & $\to$ & $1.2$ & $-$ & $133$ & $512$ & $0.26$\\
    \hline
    \end{tabular}
    \caption{Comparison of model parameters for various studies on active nematic turbulence with the same or similar (quasi-2D) model. Shown are the microscopic Reynolds number $\mathrm{Re}_\mathrm{n}$, the Ericksen number $\mathrm{Er}$, the active number $\mathrm{R}_\mathrm{a}$, the friction number $\mathrm{R}_\mathrm{f}$ as well as the box length $L$, the number of grid points $N_x$ and the grid spacing $\Delta x$. While most studies feature a larger parameter regime than displayed here, we chose for easier comparison only a subset with values from the main results.}
    \label{tab:para_comparison}
\end{table*}{}

\section{Energy and enstrophy budget}
\label{app:budg_def}
The energy budget is the time evolution of the kinetic energy spectrum defined as
\begin{align}
    E(k, t) = \frac{1}{2\Delta k} \sum_{k\le|\boldsymbol{k}|<k+\Delta k} |\hat{\boldsymbol{u}}(\boldsymbol{k}, t)|^2\,\text{,}
\end{align}
where $\boldsymbol{k}$ is the wave vector, $\Delta k=2\pi/L$, and the hat indicates the Fourier-transformed field. Contributions to the budget can be obtained term by term from the Navier-Stokes equation when taking the time derivative of the energy spectrum:
\begin{align}
    \partial_t\,E(k,t) &= \frac{1}{\Delta k} \sum_{k\le|\boldsymbol{k}|<k+\Delta k} \Re\left[ \hat{\boldsymbol{u}}^*\cdot\partial_t \hat{\boldsymbol{u}}\right]\\
    &= T + P + D + S_\mathrm{e} + S_\mathrm{a} + F\,\text{,}
\end{align}
where $\Re\left[\cdot\right]$ denotes the real part of the complex fields and the star denotes the complex conjugate. The energy transfer $T(k, t)$ originates from the self-advection of the velocity and is responsible for a transfer of energy between the scales:
\begin{align}
    T(k,t)= \frac{1}{\Delta k} \sum_{k\le|\boldsymbol{k}|<k+\Delta k} \Re\left[ \hat{\boldsymbol{u}}^* \cdot (-i \boldsymbol{k} \cdot\widehat{\boldsymbol{u}\boldsymbol{u}}) \right]\,\text{.}
\end{align}
Its sum over all wavenumbers is zero~\cite[p.~446]{davidson2015turbulence}. The pressure contribution $P(k, t)$ is identical to zero on all scales for an incompressible flow:
\begin{align}
    P(k, t) &= \frac{1}{\Delta k} \sum_{k\le|\boldsymbol{k}|<k+\Delta k} \Re\left[\hat{\boldsymbol{u}}^* \cdot \frac{1}{\mathrm{Re}_\mathrm{n}}(-i\boldsymbol{k}\hat{p})\right] \\
    &= \frac{1}{\Delta k} \sum_{k\le|\boldsymbol{k}|<k+\Delta k} \Re\left[\frac{1}{\mathrm{Re}_\mathrm{n}}(i\boldsymbol{k} \cdot \hat{\boldsymbol{u}})^* \hat{p}\right] \stackrel{\text{inc.}}= 0
\end{align}
Viscous dissipation extracts energy across scales and favors small scales due to its quadratic dependence on the wavenumber:
\begin{align}
    D(k,t)= \frac{1}{\Delta k} \sum_{k\le|\boldsymbol{k}|<k+\Delta k} \Re\left[ \hat{\boldsymbol{u}}^* \cdot \frac{1}{\mathrm{Re}_\mathrm{n}}(- \boldsymbol{k}^2 \hat{\boldsymbol{u}}) \right]
\end{align}
The active and elastic stress contributions to the budget depend on spatial changes in the orientational field $\boldsymbol Q$:
\begin{align}
    S_\mathrm{a}(k, t) &= \frac{1}{\Delta k} \sum_{k\le|\boldsymbol{k}|<k+\Delta k} \Re\left[ \hat{\boldsymbol{u}}^* \cdot \frac{\mathrm{R}_\mathrm{a}}{\mathrm{Re}_\mathrm{n}\mathrm{Er}}(-i\boldsymbol{k}\cdot\hat{\boldsymbol{\sigma}}_\mathrm{a}) \right]\\
    S_\mathrm{e}(k, t) &= \frac{1}{\Delta k} \sum_{k\le|\boldsymbol{k}|<k+\Delta k} \Re\left[ \hat{\boldsymbol{u}}^* \cdot \frac{1}{\mathrm{Re}_\mathrm{n}\mathrm{Er}}(i\boldsymbol{k}\cdot\hat{\boldsymbol{\sigma}}_\mathrm{e}) \right]
\end{align}
Linear friction extracts energy proportional to the energy spectrum, i.e. primarily at scales of high energy:
\begin{align}
    F(k,t)= \frac{1}{\Delta k} \sum_{k\le|\boldsymbol{k}|<k+\Delta k} \Re\left[ \hat{\boldsymbol{u}}^* \cdot \frac{\mathrm{R}_\mathrm{f}}{ \mathrm{Re}_\mathrm{n}}(-\hat{\boldsymbol{u}}) \right]
\end{align}{}
\begin{figure*}
    \centering
    \includegraphics[width=\textwidth]{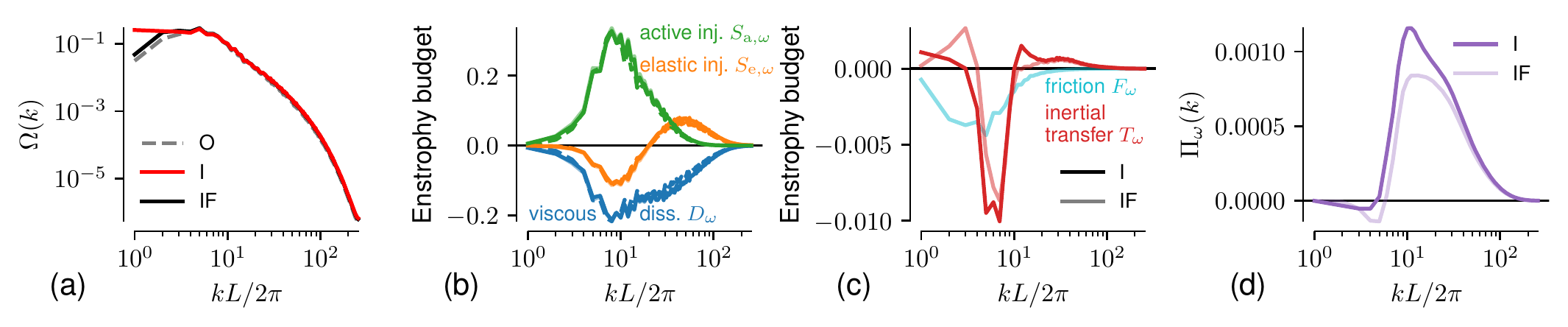}
    \caption{Advective inertia mediates a direct enstrophy transfer to the smaller scales. The original (O, dashed) regime is compared to the inertial (I, solid, dark) regime as well as to the regime with inertia and friction (IF, solid, light): ensemble- and time-averaged (a) enstrophy spectrum $\Omega(k)$, (b) and (c) enstrophy budget contributions, (d) enstrophy flux $\Pi_{\omega}(k)$. ($\mathrm{R}_\mathrm{a}=0.2, L=204.8$)}
    \label{fig:figure_2_app}
\end{figure*}
To complement the analysis of the energy budget, we also include an analysis of the enstrophy budget. The enstrophy spectrum is defined as:
\begin{align}
    \Omega(k, t) = \frac{1}{2\Delta k} \sum_{k\le|\boldsymbol{k}|<k+\Delta k} |\hat{\omega}(\boldsymbol{k}, t)|^2 \approx k^2 E(k,t)\,\text{,}
\end{align}
where the pseudo-scalar $\omega=\left\{\nabla\times\boldsymbol{u}\right\}_z$ is the vorticity. It is approximately proportional to the energy spectrum in the discrete case where the sum is over $k\le|\boldsymbol k|<k+\Delta k$ and exactly proportional in the continuous case where the sum is over $k=|\boldsymbol k|$.
The enstrophy budget is defined analogously by differentiating the enstrophy spectrum with respect to time. Compared to the energy budget, it emphasizes effects on smaller scales (higher wavenumbers) since it is effectively the energy budget multiplied with the wavenumber squared. For comparison, we have also evaluated the enstrophy budget from our simulations (Fig.~\ref{fig:figure_2_app}). The enstrophy budget nicely demonstrates that the elastic stress also injects enstrophy (as well as energy) on smaller scales. It shows a direct enstrophy transfer towards smaller scales, which is similar to classical 2D turbulence. On top of that, it features a smaller inverse transfer of enstrophy towards larger scales.

\section{Comparison of parameter regimes}
\begin{figure*}
    \centering
    \includegraphics[width=\textwidth]{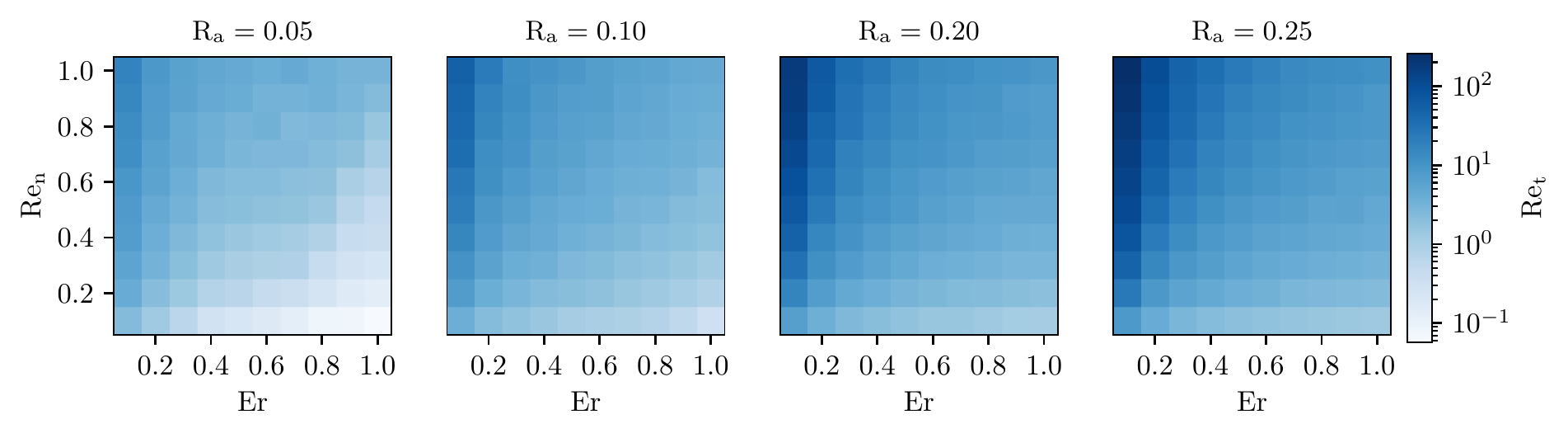}
    \caption{The importance of inertia, quantified by the turbulent Reynolds numbers $\mathrm{Re}_\mathrm{t}$, increases directly with the microscopic Reynolds number $\mathrm{Re}_\mathrm{n}$ and inversely with the Ericksen number. The simulations were performed in the regime with inertia and friction and were averaged over $30$ independent random initial conditions. ($\mathrm{R}_\mathrm{f}=7.5\cdot10^{-4}, L=204.8$)}
    \label{fig:figure_3_app}
\end{figure*}

The non-dimensionalization discussed in the sections above features, next to the active number $\mathrm{R}_\mathrm{a}$, the microscopic Reynolds and Ericksen numbers, $\mathrm{Re}_\mathrm{n}$ and $\mathrm{Er}$, respectively, which here depend on the nematic scales of the system. We scan a range of values for $\mathrm{Re}_\mathrm{n}$ and $\mathrm{Er}$ for fixed $\mathrm{R}_\mathrm{a}$ and measure the turbulent Reynolds number $\mathrm{Re}_\mathrm{t}$ (Fig.~\ref{fig:figure_3_app}), which provides a typical measure for the strength of inertial forces in turbulent flows. Extending the information of Fig.~4 in the main text, Fig.~\ref{fig:figure_3_app} includes parameter scans for two more values of the active number $\mathrm{R}_\mathrm{a}$.

For all activities, the turbulent Reynolds number increases with increasing $\mathrm{Re}_\mathrm{n}$ and decreasing Ericksen number. Depending on the active number $\mathrm{Ra}$ it takes values larger than unity even for small microscopic Reynolds numbers.

%

\end{document}